# A Deep Learning Framework for Evaluating Dynamic Network Generative Models and Anomaly Detection

Alireza Rashnu, Sadegh Aliakbary*

*Faculty of Computer Science and Engineering, Shahid Beheshti University, Tehran, Iran*

ABSTRACT

Understanding dynamic systems like disease outbreaks, social influence, and information diffusion requires effective modeling of complex networks. Traditional evaluation methods for static networks often fall short when applied to temporal networks. This paper introduces DGSP-GCN (Dynamic Graph Similarity Prediction based on Graph Convolutional Network), a deep learning-based framework that integrates graph convolutional networks with dynamic graph signal processing techniques to provide a unified solution for evaluating generative models and detecting anomalies in dynamic networks. DGSP-GCN assesses how well a generated network snapshot matches the expected temporal evolution, incorporating an attention mechanism to improve embedding quality and capture dynamic structural changes. The approach was tested on five real-world datasets: WikiMath, Chickenpox, PedalMe, MontevideoBus, and MetraLa. Results show that DGSP-GCN outperforms baseline methods, such as time series regression and random similarity assignment, achieving the lowest error rates (MSE of 0.0645, MAE of 0.1781, RMSE of 0.2507). These findings highlight DGSP-GCN's effectiveness in evaluating and detecting anomalies in dynamic networks, offering valuable insights for network evolution and anomaly detection research.

## 1. Introduction

Complex network structures are ubiquitous in various real-world systems, ranging from social networks to biological systems and technological infrastructures [1]. Studying these networks has become increasingly important in recent years as they provide opportunities to understand the behavior and dynamics of complex systems [2]. In particular, the analysis of complex networks has been used to gain insights into a wide range of phenomena, including the spread of diseases [3] and the diffusion of information [4]. Network generation models are a powerful tool for understanding, analyzing, simulating, and designing complex systems that can be represented as networks. Network modeling plays a crucial role in helping us understand the intricate structure and organization of interconnected systems with the aim of understanding how the system functions and responds to different perturbations. By modeling the dynamics of complex systems, we can simulate and analyze how information, influence, or phenomena spread through the network like epidemics [5].

On the other hand, evaluating the output of network generative models can be difficult because there is no clear objective measure of what constitutes a "good" output. Unlike discriminative models, where the output can be evaluated based on its accuracy in predicting a known label or class, generative models are designed to create new data similar to the training data. This means that any specific criteria do not necessarily constrain the output of a generative model and can be highly subjective. Furthermore, generative models often produce probabilistic outputs, meaning that the same input can result in different outputs each time the model is executed. This makes it difficult to compare the output of a generative model to a ground truth dataset, as there may be multiple valid outputs for a given input. As a result, evaluating the output of generative models often requires a combination of quantitative and qualitative analysis and human judgment. Generally, researchers have used three approaches to evaluate the output of network generative models:

1) The structural features of an artificial graph (such as the degree distribution, the clustering coefficient distribution, and transitivity) and its real counterpart are compared [6-8].
2) Indirect assessment. A classification model is trained with real graphs and tested with generated graphs. If the artificial graph is similar to the target graph, the classification model gives a score of one; otherwise, a zero score is received [9].
3) Quality-based approach. The identical edges in the structure of the generated graph and the real graph are kept constant for model evaluation. In contrast, the other links of the synthetic graph nodes are changed randomly. In this case, if the statistical parameters of the synthetic graph, such as degree distribution, density, diameter, etc., do not change compared to the real graph, the generating model has shown good performance [10].

* *Corresponding author.* Tel.: +98-21-29904110
E-mail addresses: a.rashnou@alumni.sbu.ac.ir (A. Rashnu), s_aliakbary@sbu.ac.ir (S. Aliakbary)

Although these methods are inherently designed for evaluating static graph generative models, some dynamic generative models have used them for model evaluation [11-15]. Dynamic networks change; therefore, evaluation methods designed for static networks may not be suitable. For example, metrics that measure the centrality of nodes in a static network may not be useful for understanding the dynamics of a network over time because it will not necessarily have a fixed value.

Temporal networks, characterized by their evolving connections over time, introduce a layer of complexity beyond traditional static network models. Predicting node status within such dynamic contexts necessitates a nuanced understanding of how graph similarity, often explored in static settings, translates to temporal dynamics. Our research addresses this critical gap by elucidating the interconnectedness between graph similarity metrics and the evolving states of nodes in temporal networks. By leveraging insights from graph similarity learning, we discern patterns in temporal network dynamics that influence node status predictions. Our approach acknowledges the dynamic nature of real-world systems, where nodes interact and evolve over time, rendering traditional static analyses insufficient for capturing the full spectrum of network behaviors. Furthermore, our work recognizes the limitations of existing evaluation methods designed primarily for static graph generative models when applied to dynamic network settings.

In this paper, we consider the challenge of evaluating dynamic complex network generative models' output using different graph embedding mechanisms, recurrent neural networks (RNN), and fully connected layers. In other words, given the history of a dynamic network and a new snapshot, the proposed model called DGSP-GCN (Dynamic Graph Similarity Prediction based on Graph Convolutional Network) predicts how likely the hypothetical snapshot will be the future of the same temporal network. While it is true that DGSP-GCN leverages existing embedding methods, its contribution lies in the novel synthesis, customization, and application of these techniques within a unified framework tailored for dynamic graph node-level similarity prediction. Our experiments illustrate that the proposed model outperforms the baselines. The main contributions of this paper are as follows:

1) A Deep Learning-Based Evaluation Method for Dynamic Network Generative Models: We introduce DGSP-GCN, a graph convolutional network-based approach that effectively captures both spatial and temporal dependencies in dynamic graphs. This enables a robust and accurate assessment of the quality of generated dynamic networks;
2) A Unified Framework for Anomaly Detection in Temporal Complex Networks: Our proposed method not only evaluates the fidelity of synthetic networks but also detects anomalies in real-world temporal graphs by leveraging dynamic graph signal processing techniques. This provides a versatile and effective tool for analyzing evolving network structures;
3) Empirical Validation on Real and Synthetic Datasets: We evaluate DGSP-GCN on multiple benchmark datasets, demonstrating its superiority in both evaluating generative models and detecting anomalies in dynamic graphs. Our results confirm that DGSP-GCN outperforms existing baselines in predictive accuracy and robustness.

The rest of this paper is organized as follows: Section 2 reviews the state-of-the-art graph similarity prediction models. In Section 3, the problem statement is presented. Section 4 illustrates our proposed method. Section 5 shows the experimental evaluations. Finally, section 6 concludes and explains the future works.

## 2. Literature review

In the vast realm of data analysis, understanding the unique attributes and relationships within complex structures has appeared as a paramount challenge. Within this context, graph similarity learning has emerged as an intriguing avenue, enabling researchers to uncover hidden correlations, discover underlying patterns, and extract valuable insights from interconnected data. The primary goal of graph similarity learning is to develop effective techniques that capture the inherent similarities and dissimilarities between graphs [16]. We can discern their structural, topological, and semantic characteristics by measuring the similarity between graphs. This holistic understanding allows us to categorize graphs more accurately, identify anomalies, and better understand their underlying dynamics [17]. For instance, in social network analysis, graph similarity learning can help identify communities or clusters of individuals with similar social connections. Moreover, graph similarity learning has applications in recommendation systems, which can be used to identify similar users or items based on their interconnected relations.

In summary, graph similarity learning is a vital tool in data analysis, allowing us to unlock hidden insights, understand complex structures, and make informed decisions in various domains. Our work aims to reconcile the realms of graph similarity learning with the intricacies of temporal network dynamics, shedding light on evolving system states and facilitating predictive insights into node behaviors. Generally, graph similarity learning approaches are divided into categories, including graph kernels, graph embedding methods, and graph neural networks (GNN). We will examine each one below.

### *2.1. Methods based on graph kernels*

A graph kernel is a function that measures the similarity between two graphs by mapping them into a high-dimensional feature space. Graph kernels are commonly used in machine-learning tasks involving graph-structured data [18]. The basic idea behind graph kernels is to define a function that maps each graph into a vector of features that capture its structural properties. The similarity between the two graphs can then be computed as the inner product of their feature vectors in the high-dimensional space [19].

There are many different types of graph kernels, each with its strengths and weaknesses. Some popular graph kernels include the random walk kernel [20], the subtree kernel [21], and the neighborhood hash kernel [22]. The choice of kernel depends on the specific application and the properties of the graphs being analyzed. While graph kernel methods have many advantages, they also face several challenges that must be carefully considered when applying them to real-world problems [16]. Here are some of the main challenges:

- Computational Complexity: Graph kernel methods can be computationally expensive, especially for large graphs. Since these methods involve comparing graphs based on structural or topological properties, the computations can become time-consuming and resource-intensive as the size of the graphs increases. This can limit their scalability and efficiency in handling large-scale graph datasets.
- Kernel Choosing: There are many different types of graph kernels, each with its own strengths and weaknesses. Choosing the right kernel for a particular problem can be challenging, and there is often no clear best choice.
- Sensitivity to Graph Representations: Graph kernel methods heavily rely on the representations of graphs, such as node or edge labels, that are provided as input. Small changes or variations in these representations may lead to significantly different kernel values, affecting the similarity measures between graphs.

### *2.2. Graph embedding methods*

Graph embedding methods for similarity are techniques used to represent graphs as low-dimensional vectors, which can be used to measure similarity between graphs. These methods aim to capture the structural and semantic information of the graph in the embedding space, such that similar graphs are mapped to nearby points in the embedding space. There are various graph embedding methods for similarity, including node and graph embedding methods. In the case of node embedding, the aim is a representation of each node to a vector by some methods like node2vec [23, 24], which can be aggregated to obtain an embedding for the entire graph [25]. Graph embedding methods aim to directly learn the representation of the entire graph by considering the graph structure like [26-28]. However, there are several challenges associated with graph embedding methods, including [17]:

- Heterogeneity: Graphs can be heterogeneous, containing different nodes and edges. Embedding methods need to handle this heterogeneity and capture the relationships between different types of nodes and edges.
- Structure-oriented: Although structural features such as node degree distribution, clustering coefficient distribution, number of triangles, network diameter, etc., are used to generate vectors at the node and graph levels, the node and edge level features are not considered for embedding.
- Loss of Graph Structure Interpretability: Embedding methods aim to represent graphs in low-dimensional vector spaces. While this enables numerical comparisons and similarity metrics, it can lead to a loss of interpretability in terms of the original graph structure. The transformed representations may not directly reveal the inherent graph properties and relationships, making comprehending the reasons behind similarity or dissimilarity scores challenging.

### *2.3. GNN-based methods*

GNN methods are a class of machine learning techniques that have emerged as powerful tools for graph similarity prediction. By leveraging their ability to capture and learn from complex graph structures, GNNs offer a promising approach for comparing the similarity of different graphs. Through a series of iterative aggregation and transformation steps, GNNs can effectively encode the inherent structural properties of graphs into low-dimensional representations, commonly referred to as node or graph embeddings [29-31]. Not only do these learned embeddings encapsulate the topological relationships and attributes of individual nodes, but they also capture the global structural patterns and dependencies present in the graph as a whole. By harnessing the expressive power of GNNs, graph similarity prediction can benefit from the rich representations learned by the network, facilitating more accurate and nuanced comparisons between complex and heterogeneous graph structures in diverse domains.

One popular approach for graph similarity prediction using GNNs is to use Siamese networks [32-35], which consist of two identical GNNs that take in two different graphs as input and output a similarity score. The two GNNs share the same weights, allowing them to learn a common representation of the graphs. Another approach is to use a contrastive loss function, which encourages the GNN to learn representations that are close together for similar graphs and far apart for dissimilar graphs [36]. This can be combined with a Siamese network architecture to learn a similarity function. Other GNN-based approaches for graph similarity prediction include using attention mechanisms to focus on important substructures within the graphs [37]. While Graph Neural Network (GNN) based methods have shown promising results in graph similarity learning, they also have a few disadvantages.

Here are some of them [17]:

- Computational Complexity: GNNs can be computationally expensive, especially for large graphs with a high number of nodes and edges. The complexity increases as the graphs' size and complexity grow, making it challenging to scale GNN-based methods to large-scale graph similarity learning tasks.
- Interpretability and Explainability: The complex nature of the GNN architecture makes it challenging to understand how and why certain patterns are learned and used for similarity comparisons. Interpreting the decisions made by GNN-based models can be difficult.

## 3. Proposed method

### *3.1. Problem statement*

With the help of synthesized networks, we can represent complex systems through graph structure. The node's connections in real networks are a specific and meaningful pattern. Therefore, the corresponding synthesized network should match the real network. Put differently, the closer the synthetic network is to the target network, the more precise the outcomes of different tests conducted on the synthetic networks will be.

If $\mathbb{G}$ is a dynamic complex network and its snapshots contain $\{G_1, G_2,\dots, G_T\}$, then the problem is to predict the similarity of a network $G_r$ (perhaps a synthesized graph) with $G_{T+1}$ of $\mathbb{G}$. To formalize this prediction task, Eq. (1) introduces the inputs and output of the problem, where $f$ is a function that takes in the sequence $\{G_1, G_2,\dots, G_T\}$ and $G_r$ to compute the similarity and $S(G_{T+1},G_r)$ is the similarity between $G_{T+1}$and $G_r$. We assume that the considered networks are static graph-temporal signals. This implies that the arrangement of the network remains constant throughout time, but the attributes of the network nodes alter over time.

$$f(\{G_1, G_2, \dots, G_T\}, G_r) = S(G_{T+1}, G_r) \tag{1}$$

One of the paramount applications of predicting the similarity between evolving network states lies in anomaly detection within dynamic complex networks. Sudden deviations in the similarity score of $f(\{G_1, G_2, \dots, G_T\}, G_{T+1})$ might indicate potential anomalies, such as malicious activities or unexpected patterns. Consequently, leveraging this similarity-based approach offers a proactive mechanism to identify and mitigate threats or disruptions in dynamic network environments. Furthermore, this predictive framework is pivotal in evaluating the efficacy and performance of dynamic generative models. Generative models that emulate and reproduce complex networks' structural and temporal characteristics necessitate rigorous evaluation metrics. Researchers and practitioners can quantitatively assess dynamic generative models' fidelity, robustness, and generalization capabilities by juxtaposing the predicted similarity scores with ground truth or benchmark snapshots. Such evaluations ensure that generative models capture essential temporal dynamics, structural nuances, and emergent behaviors inherent to real-world complex networks, thereby fostering advancements in network synthesis, simulation, and reconstruction methodologies.

**Algorithm 1.** Algorithm of noise injection approach

**Input:** Buckets # $\mathbf{Bucket}_i$={$\mathbf{G_0}$, $\mathbf{G_2}$,…, $\mathbf{G}_T$, $\mathbf{G}_r$}
**Output:** List of labels, Buckets

```
Dictionary ←{}
List_of_labels[len(Buckets)] ←{1}
Number_of_nodes ← len(Buckets[0][0].nodes)
for i=0 to Number_of_nodes do
    Dictionary[i] ←[0.0]
for bucket : Buckets do
    for node = 0 to Number_of_nodes do
        if min(bucket[node]) < Dictionary[node][0] then
            Dictionary[node][0] ← min(bucket[node])
        if max(bucket[node]) > Dictionary[node][1] then
            Dictionary[node][1] ← max(bucket[node])
Index_of_randomly_selected_buckets ← random(range(0, len(Buckets) - 1), random(range(0, len(Buckets) - 1), 1))
for bucket_index : Index_of_randomly_selected_buckets do
    Number_of_randomly_selected_nodes ← random(range(0, len(Buckets[0][0].nodes) - 1), 1)
    Index_of_randomly_selected_nodes ← random(range(0, len(Buckets[0][0].nodes)-1), Number_of_randomly_selected_nodes)
    List_of_labels[Bucket_index] ← 1− (Number_of_randomly_selected_nodes / (len(Buckets[0][0].nodes)))
    for node_index :  Index_of_randomly_selected_nodes do
        Buckets[bucket_index][node_index].node_feature[-1] ← random(Dictionary[node_index][0], Dictionary[node_index][1])
```

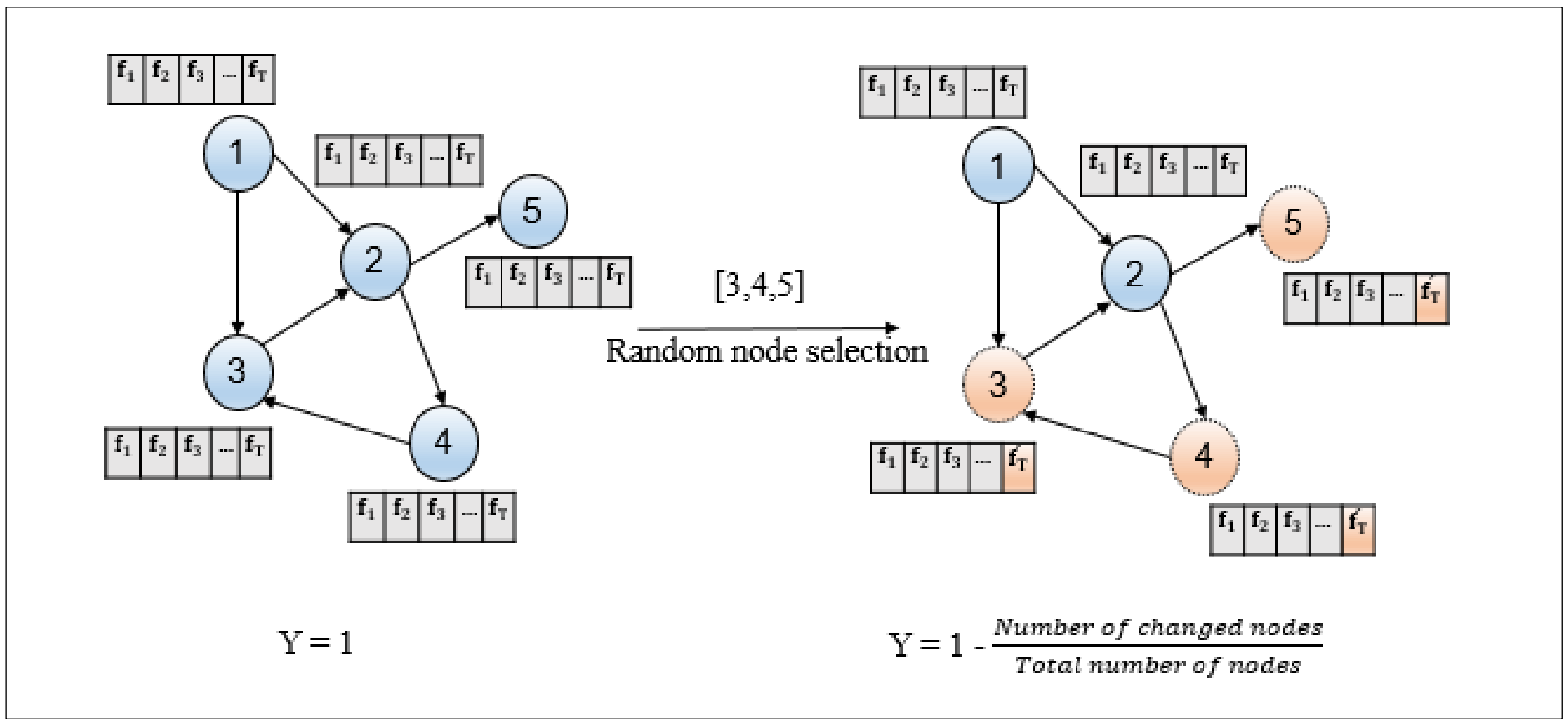

**Fig. 1.** The process of preparing datasets with the help of the noise injection approach to train the proposed model.

### *3.2. Noise injection approach*

Referring to Eq. (1), when $G_r$ aligns perfectly with $G_{T+1}$, the resultant similarity metric will be unity (i.e., S ($G_{T+1}$,$G_r$) = 1). Conversely, any divergence or alteration in $G_r$ leads to a proportional decrement in the similarity value. To illustrate, if $G_r$ undergoes a 20% modification, the similarity is quantified as S ($G_{T+1}$,$G_r$) = 0.8. Motivated by this foundational understanding, we harness the concept of noise injection to curate a comprehensive training dataset for our deep learning-based model, denoted as *f*. This research's datasets encompass distinct temporal snapshots organized into various buckets. We systematically introduce varied noise levels into $G_r$ within these buckets, representing the terminal snapshot. Subsequently, we delineate the label for each bucket predicated on the computed similarity distance, thereby facilitating a robust training paradigm for our predictive model. Fig. 1 and Algorithm 1 illustrate the noise injection approach. First of all, we assign a Y = 1 similarity label to each of the buckets of datasets, which shows the degree of complete similarity of the last snapshot with its real state. Then, we randomly select several buckets. Next, some nodes are randomly selected, like nodes 3, 4, and 5 in Fig.1, and to inject a logical noise, a random value between the minimum and maximum value that the node has in the entire dataset is replaced by the last feature of the node. Finally, the bucket similarity label is calculated using Eq. (2).

$$Y_{Bucket[i]} = 1 - \frac{Number\ of\ changed\ nodes}{Total\ number\ of\ nodes} \tag{2}$$

### *3.3. DGSP-GCN method*

In this study, we introduce DGSP-GCN, a novel deep learning-based framework that integrates dynamic graph signal processing (DGSP) techniques with GCNs for the evaluation of generative models and anomaly detection in temporal complex networks. The datasets used in this research are dynamic complex networks, requiring the integration of GNNs and RNNs to effectively capture both spatial and temporal dependencies. The GNN layers enable the network to learn node and graph representations by propagating information across neighboring nodes, preserving both local and global structural relationships. This is achieved through an iterative message-passing mechanism, where each node aggregates information from its neighbors to update its representation.

Fig. 2 illustrates the architecture of our proposed DGSP-GCN model, which consists of four main stages:

1. Graph Embedding via GCN
   - Each snapshot of the temporal network is processed using a GCN layer to generate 32-dimensional embeddings for nodes and edges.
   - This transformation captures the structural properties of the graph while preserving connectivity information.
2. Temporal Feature Extraction via RNNs
   - The sequence of graph embeddings from multiple snapshots is passed through an RNN layer (e.g., LSTM or GRU).
   - This step models the evolution of node relationships over time, capturing temporal dependencies in the dynamic network.
3. Graph-Level Representation via Mean Pooling
   - A mean pooling layer aggregates node-level embeddings into a compact representation of the entire graph sequence.
   - This enables the model to perform similarity forecasting at the graph level.

4. Similarity Prediction via Multilayer Perceptron (MLP)
   - The pooled representation is passed through a three-layer MLP with 32, 64, and 1 neurons, respectively.
   - The final output is a predicted similarity score for the last snapshot $G_r$ relative to previous snapshots $G_1$ to $G_T$.
   - The snapshot $G_r$ corresponds to $G_{T+1}$ in the input bucket, where 50% noise injection is applied during training to enhance model robustness.

The similarity prediction process by the proposed method is outlined in Algorithm 2. This structured approach ensures that both spatial and temporal relationships within dynamic graphs are captured effectively, improving the accuracy of generative model evaluation and anomaly detection.

To develop the most effective model architecture, we conducted extensive experimental comparisons using different node embedding techniques, including:

1) GConvGRU (Graph Convolutional Gated Recurrent Unit) [38]. It consists of multiple layers of GConvGRU cells. Each cell has two main components: a GCN layer and the Gated Recurrent Unit (GRU) layer.
2) GConvLSTM (Graph Convolutional Long Short-Term Memory) [38]. It combines the GCN and long short-term memory (LSTM) networks to capture both spatial and temporal dependencies in the graph data.
3) TGCN (Temporal Graph Convolutional Network) [39]. Its architecture comprises several layers, such as GCN, GRU, and temporal pooling.
4) AGCRN (Adaptive Graph Convolutional Recurrent Network) [40]. It consists of two main components, including gated convolutional layers and recurrent units, to effectively capture spatial and temporal dependencies in the data.
5) A3TGCN (Attention Temporal Graph Convolutional Network) [41]. The A3T-GCN employs a soft attention mechanism to calculate the importance of each hidden state from the historical time points and to generate a context vector that encapsulates global temporal variations. The attention mechanism consists of the following key steps and equations:
   1. Score Calculation: A scoring function is used to compute the score $e_i$, Eq. (3), for each hidden state $h_i$ at time i. The score measures the relevance of $h_i$ to the prediction task. In the A3TGCN, a two-layer perceptron is used for this purpose:

$$e_i = w^{(2)}.ReLU\left(w^{(1)} + b^{(1)}\right) + b^{(2)} \tag{3}$$

   Here:
   - $w^{(1)}$ and $b^{(1)}$ are the weight and bias for the first layer.
   - $w^{(2)}$ and $b^{(2)}$ are the weight and bias for the second layer.
   - ReLU is the rectified Linear Unit activation function.

   2. Weight calculation: The scores $e_i$ are normalized using a Softmax function to compute the attention weights $\alpha_i$, Eq. (4). This ensures the weights sum to 1, representing the relative importance of each hidden state:

$$\alpha_i = \frac{\exp(e_i)}{\sum_{k=1}^{n}\exp(e_k)} \tag{4}$$

   Here, n is the total number of time points in the historical sequence.

   3. Context Vector: According to the Eq. (5), the context vector $C_t$ is calculated as a weighted sum of the hidden states, where the weights $\alpha_i$ represent the importance of each state:

$$\mathrm{C}_t = \sum_{i-1}^{n} \alpha_i \,.\, \mathrm{h}_i \tag{5}$$

   The context vector $C_t$ summarizes the global temporal variation information, which is then used in subsequent layers for prediction.

These equations allow the A3TGCN to dynamically adjust the influence of different time points, enabling it to better capture long-term dependencies and improve forecasting accuracy.

Through a series of controlled experiments in Section 4 (Experiments), we find that the A3T-GCN architecture consistently outperforms other alternatives in terms of both predictive accuracy and computational efficiency when integrated with dynamic graph signal processing techniques. As a result, we adopt the A3T-GCN architecture as the core of our DGSP-GCN framework, ensuring that it effectively captures evolving graph structures while maintaining computational feasibility.

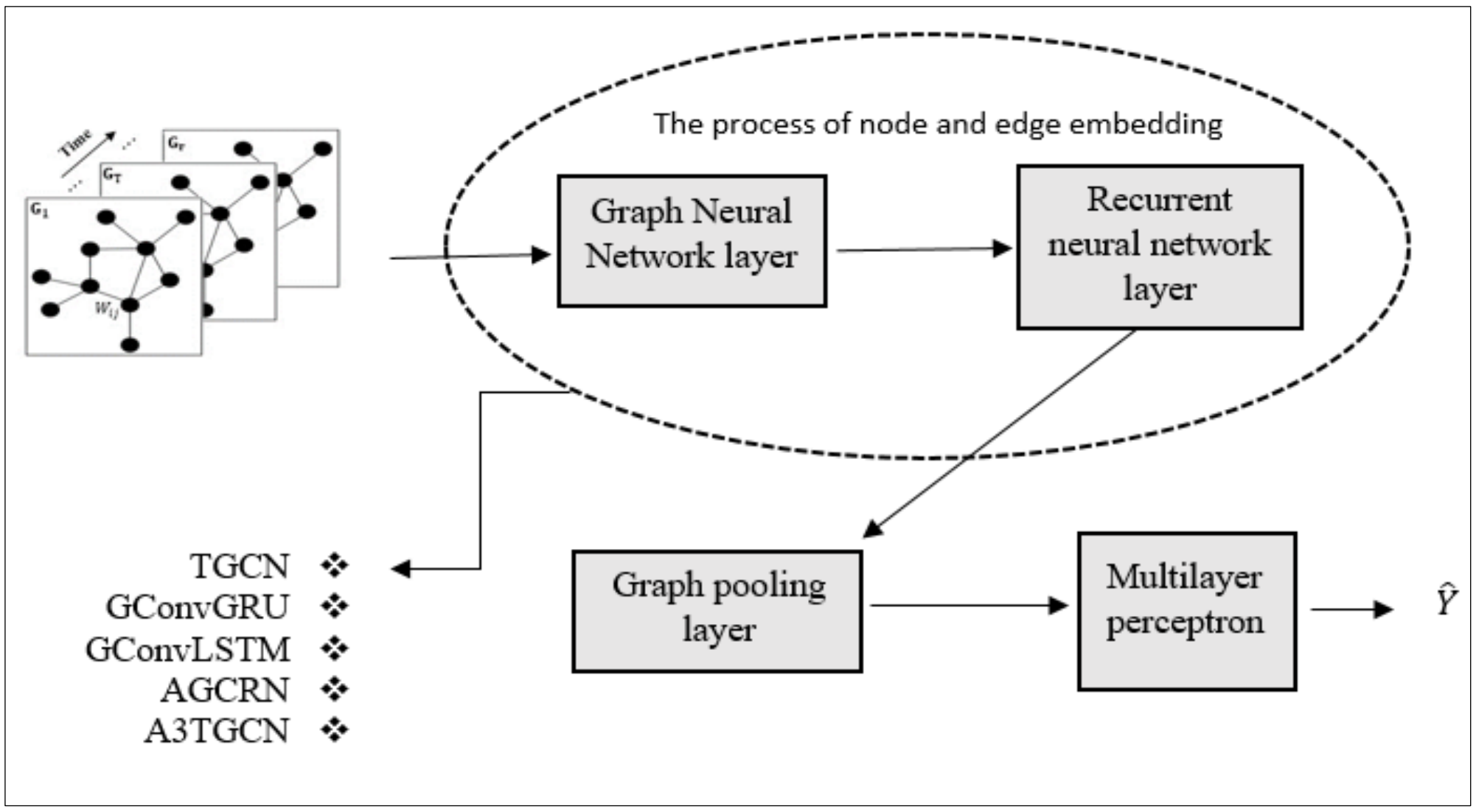

**Fig. 2.** The architecture of DGSP-GCN model

**Algorithm 2.** Algorithm of DGSP-GCN

**Input:** Bucket # $\mathbf{Bucket}_i$ = {$\mathbf{G_0}$, $\mathbf{G_2}$,…, $\mathbf{G}_T$, $\mathbf{G}_r$}

**Output:** Y_hat # (predicted similarity between $\mathbf{G}_{T+1}$ and $\mathbf{G}_r$)

```
1.  Vectors_of_nodes ← []
2.  for snapshot : Bucket.snapshots do
3.      Node_embedding_list ← []
4.      for node : snapshot.nodes do
5.          Node_embedding_list.append(graph_convolution_layer(snapshot))  # 32-dimensional vector for each node
6.      Vectors_of_nodes.append(recurrent_neural_network(Node_embedding_list) # recurrent neural network like LSTM
7.  Vector_of_bucket ← 0
8.  for vector : Vectors_of_nodes do
9.      Vector_of_bucket ←\gets← Vector_of_bucket + vector
10. Y_hat{Y} ← dense_layer_with_32neurons(Vector_of_bucket / len(Vectors_of_nodes))
11. Y_hat{Y} ← dense_layer_with_64neurons(Y_hat)
12. Y_hat{Y} ← dense_layer_with_1neuron(Y_hat{Y})
```

### 3.4. The performance measures

Evaluating machine learning models is crucial in assessing their performance and effectiveness. The choice of appropriate evaluation metrics holds immense importance due to objective assessment. In other words, evaluation metrics provide an objective and standardized way of measuring and comparing model performance. To evaluate the performance of our regression model, it is common to use Eq. (6), Eq. (7), and Eq. (8), where N is the number of samples, Y is the real label, and $\hat{\mathbf{Y}}$ is the label predicted by the model.

$$Mean\ Squared\ Error\ (MSE) = \frac{1}{N}\sum_{i=1}^{N}\left(Y - \hat{Y}\right)^2 \quad (6)$$

$$Mean\ Absolute\ Error\ (MAE) = \frac{1}{N}\sum_{i=1}^{N}\left|\left(Y - \hat{Y}\right)\right| \quad (7)$$

$$Root\ Mean\ Squared\ Error\ (RMSE) = \sqrt{\frac{1}{N}\sum_{i=1}^{N}(Y - \hat{Y})^2} \quad (8)$$

Giving higher weights to larger errors is one of the advantages of MSE, thereby indicating the importance of reducing significant deviations. Nevertheless, one disadvantage of it is that it squares the errors, which can lead to an amplification of the impact of outliers. On the other hand, although MAE is less sensitive to outliers and provides a robust measure of error, it may not fully capture the relative importance of different errors. However, RMSE combines the benefits of both MSE and MAE by calculating the square root of the average squared difference between predicted and actual values. Therefore, with the help of these measures, we can evaluate the performance of the proposed model in different aspects.

## 4. Experiments

### *4.1. Evaluation methods*

In order to comprehensively assess the performance of our model, we conducted a thorough comparative analysis against a set of baselines. This evaluation methodology allows us to gauge the effectiveness and superiority of our proposed approach in tackling the given problem. However, previous methods in graph comparison are limited to static graphs, while the datasets used in this research are temporal. That is why we have presented two baselines, including time series regression and random methods, to compare the performance of the proposed model. In the case of the random method, a random number between zero and one is generated as $\hat{Y}$ for each sample of the test dataset. Although the random method's performance is not impressive, it does assure us that our proposed model for similarity prediction is not performing worse than the least effective baseline. Another baseline idea we presented is the use of time series regression. Fig. 3 and Algorithm 3 describe the process of similarity prediction. In this case, a regression model is trained for each node in every sample of the test dataset. The model gets the features of the node from snapshots 1 to T-1 to predict its feature for the next snapshot. In other words, this model predicts the last feature for each node after receiving n-1 previous features of the node. Eventually, the amount of $\hat{Y}_{Bucket[i]}$ calculate based on Eq. (9).

$$\hat{Y}_{Bucket[i]} = \frac{1}{m}\sum_{j=1}^{m} 1 - |Y_j - \hat{Y}_j| \qquad (9)$$

Where m is the number of nodes in each sample, $Y_j$ is the real label of the node of the last snapshot and $\hat{Y}_j$ is the predicted label by the time series regression model. The absolute difference between these two values represents the amount of noise injected into the last feature of each bucket node.

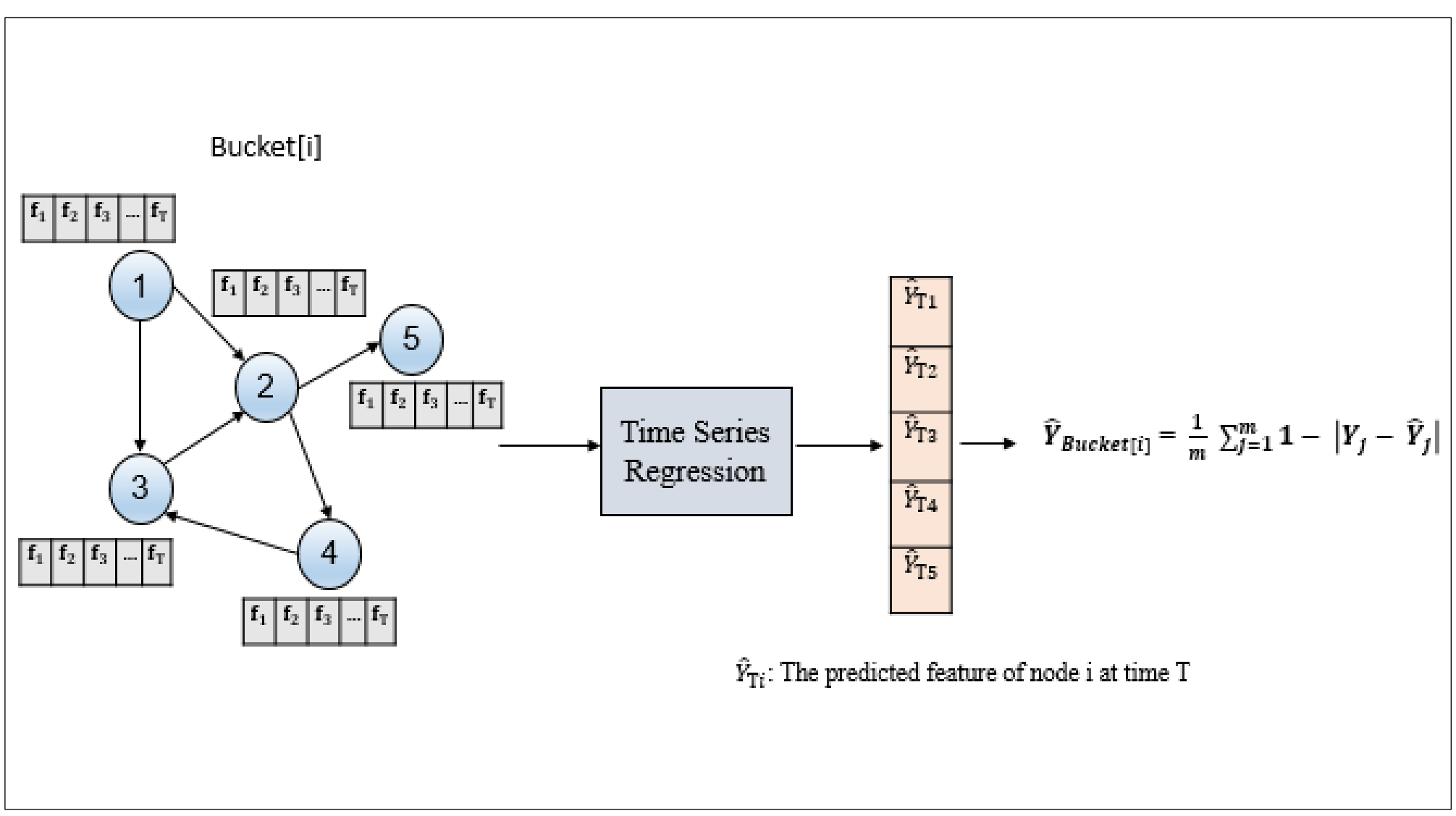

**Fig. 3.** The process of time series regression for prediction of bucket's label for each sample of the test dataset.

**Algorithm 3.** Algorithm of time series regression method

**Input:** Buckets

**Output:** Y_hat

```
Time ← []
for i : list(range(len(Buckets[0].node[0].feature) - 1)) do
    Time.append(i)
Y_hat ← []
for bucket : Buckets do
    List_Y_hat_nodes ← []
    List_Y_test_nodes ← []
    for node_features : range(len(bucket.node)) do
        node_features ← min_max_normalization(node_features)
        LinearRegression.fit(Time, node_features[0:-1])
        List_Y_hat_nodes.append(LinearRegression.predict(Time[-1]+1))
        List_Y_test_nodes.append(node_features[-1])
        Noise_list ← []
        for j : range(len(List_Y_test_nodes)) do
            Noise_list.append(1 - absolute(List_Y_test_nodes[j] - List_Y_hat_nodes[j]))
    Y_hat.append(mean(Noise_list))
```

### *4.2. Data description*

This section emphasizes a robust and comprehensive data description to provide a solid foundation for our research findings and analysis. Data plays a crucial role in shaping the outcomes of any study, and by thoroughly understanding the datasets used, we can ensure the validity and reliability of our results. Therefore, we use five real-world available datasets, which are a kind of static graph-temporal signals. They include the following:

1) WikiMath [42]. This is a collection of important math articles from Wikipedia, presented as a graph where each page is a vertex and links between them are edges. The weight of each edge represents the number of links from the source page to the target page. The target is the number of daily visits to these pages.
2) Chickenpox [43]. This is a collection of information about chickenpox cases in Hungary. The data includes the number of chickenpox cases each week, where each city is a vertex and the road between them an edge.
3) PedalMe [44]. This is a dataset of Bicycle deliveries in London. The data is represented as a graph, where different areas are the vertices, and the connections between them are the edges. The vertex features show the number of deliveries requested each week.
4) MetraLa [45]. This dataset predicts traffic patterns in the Los Angeles Metropolitan area. The data was gathered from 207 loop detectors on highways throughout Los Angeles County.
5) MontevideoBus [46]. This dataset contains information about the number of passengers who boarded buses at various stops in Montevideo city. The weight of these connections represents the distance between stops.

These datasets are summarized in Table 1. To train and evaluate our proposed model for each dataset, we use the cross-validation method with K=3.

**Table 1**: The used datasets in our experiments.

| Dataset | #Nodes | #Edges | #Snapshots | Frequently |
|---|---|---|---|---|
| **WikiMath** | 1068 | 27079 | 731 | Daily |
| **Chickenpox** | 20 | 102 | 520 | Weekly |
| **PedalMe** | 15 | 225 | 30 | Weekly |
| **MetraLa** | 207 | 1722 | 3224 | 5-Minutes |
| **MontevideoBus** | 678 | 690 | 734 | 1-Hours |

*4.3. Experimental result*

The hyperparameters used in our experiments were set as follows: the number of epochs was 30, the number of snapshots per bucket was 10, the node embedding dimension was 32, and the learning rate was 0.01. The experiments were conducted on five real-world datasets: WikiMath, Chickenpox, PedalMe, MontevideoBus, and MetraLa.

To ensure the robustness and optimal performance of the proposed DGSP-GCN model, we conducted extensive ablation studies to evaluate the impact of key hyperparameters, including the number of epochs, snapshot size per bucket, node embedding dimension, and learning rate. These studies were performed across all datasets to identify the most effective configuration for the model. For instance, we tested embedding dimensions of 16, 32, and 64 and found that a dimension of 32 consistently provided the best balance between computational efficiency and predictive accuracy. Similarly, we evaluated snapshot sizes of 5, 10, and 15 per bucket and determined that a size of 10 yielded the most stable and accurate results. The learning rate was tuned within the range of 0.001 to 0.1, with 0.01 emerging as the optimal value for minimizing error rates. Some results from these erosion studies, such as the number of snapshots per bucket and the number of epochs for the Chickenpox dataset, are presented in Fig. 4. This figure offers a clear justification for the chosen hyperparameters. This systematic approach ensures that the model's performance is not only reproducible but also optimized for the given tasks.

The results of the proposed DGSP-GCN model and baseline methods, including the random method and time series regression, are summarized in Table 2. The performance metrics used for evaluation include Mean Squared Error (MSE), Mean Absolute Error (MAE), and Root Mean Squared Error (RMSE).

Table 2 presents the prediction errors for each dataset using different embedding layers within the DGSP-GCN model, as well as the baseline methods. The A3TGCN embedding layer consistently achieves the lowest error rates across all datasets. The error rates for the random method and time series regression method are higher than those of the DGSP-GCN model. The final prediction results of the DGSP-GCN model with the A3TGCN embedding layer are detailed in Table 3. The average MSE, MAE, and RMSE across all datasets are 0.0645, 0.1781, and 0.2507, respectively. These values indicate a lower error rate compared to the baseline methods.

The bar graphs from Fig. 5 show the error rate of the proposed model and baselines based on the results of experiments in Table 2. Let's discuss the bar graphs related to the dataset of WikiMath as an example; the supplied bar chart denotes the percentage of error rates of the proposed model and other baselines based on MSE, MAE, and RMSE performance measures. As an overall trend, the lowest error rates can be observed for the proposed model with the A3TGCN layer. In contrast, these figures are higher for time series regression and especially for the random method than the others. To begin with, in MSE, the error rate for the proposed model with embedding layer including GConvGRU, GConvLSTM, TGCN, AGCRN, and A3TGCN is on an average of well over 10%. Also, the average MAE of the proposed model with different embedding architectures is almost 28%, and for RMSE, it is almost 34%. In the case of time series regression, these figures' percentages are 7%, 7%, and 8% higher than their counterparts in the proposed model with the A3TGCN embedding layer, respectively. It can also be seen that there is an almost similar trend for the other baseline.

Finally, after determining the optimal parameters and choosing an architecture for embedding, the final results of the proposed method can be seen in Table 3 with three dense layers, including 32, 64, and 1 neurons, respectively. We have used PyTorch [47] and PyTorch Geometric Temporal [48] libraries to implement the proposed model.

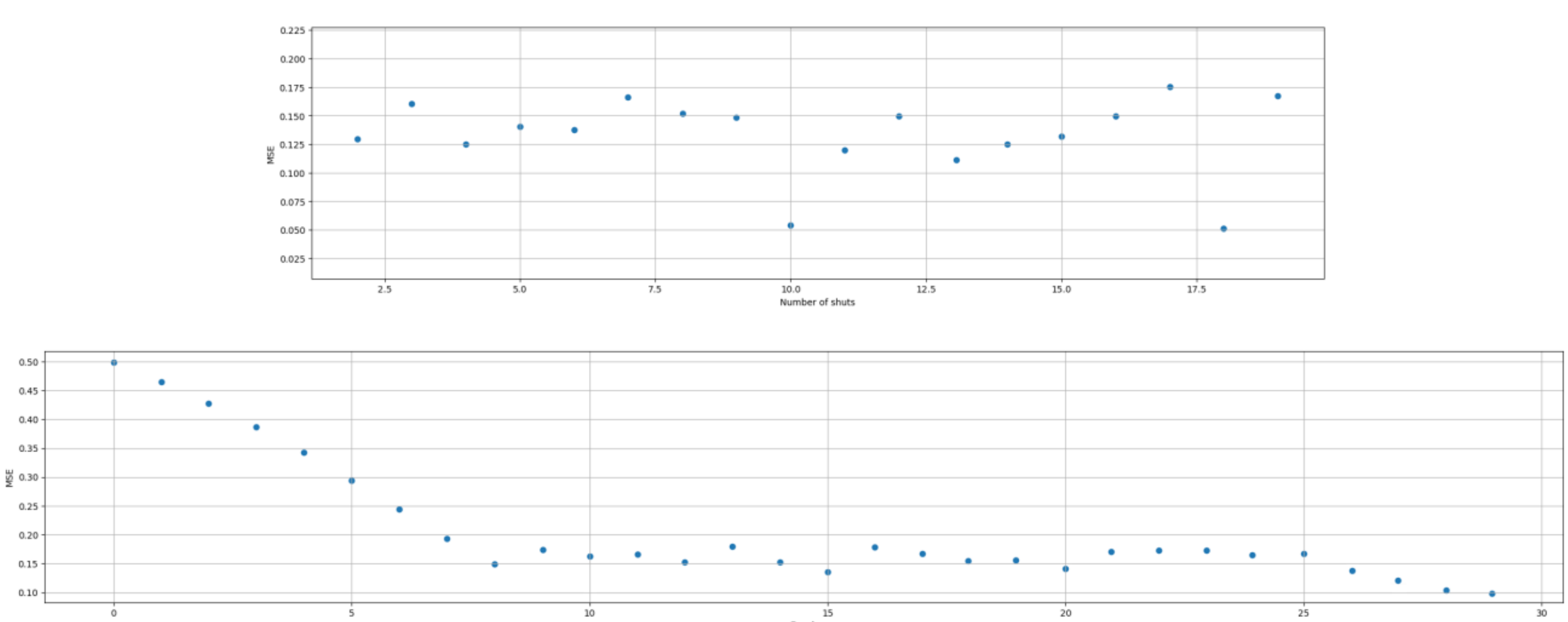

**Fig 4.** Model error on the Chickenpox dataset: The impact of the number of shots per bucket and the number of periods on MSE

**Table 2**: The prediction results of the proposed model and other baselines

| Dataset | Method | Recurrent layer | MSE | MAE | RMSE |
|---|---|---|---|---|---|
| **Wiki Math** | DGSP-GCN Method | GConvGRU | 0.1367 | 0.3264 | 0.3598 |
| | | GConvLSTM | 0.1262 | 0.3091 | 0.3579 |
| | | TGCN | 0.1117 | 0.2800 | 0.3419 |
| | | AGCRN | 0.1220 | 0.2843 | 0.3490 |
| | | A3TGCN | 0.1012 | 0.2624 | 0.3224 |
| | Random Method | - | 0.1991 | 0.3676 | 0.4460 |
| | Time series regression Method | - | 0.1724 | 0.3523 | 0.4139 |
| **Chickenpox** | DGSP-GCN Method | GConvGRU | 0.1162 | 0.3015 | 0.3442 |
| | | GConvLSTM | 0.1065 | 0.2683 | 0.3307 |
| | | TGCN | 0.1067 | 0.2697 | 0.3173 |
| | | AGCRN | 0.1043 | 0.2351 | 0.3011 |
| | | A3TGCN | 0.0834 | 0.1797 | 0.2648 |
| | Random Method | - | 0.2136 | 0.3789 | 0.4620 |
| | Time series regression Method | - | 0.1489 | 0.3273 | 0.3858 |
| **PedalMe** | DGSP-GCN Method | GConvGRU | 0.0904 | 0.3120 | 0.3016 |
| | | GConvLSTM | 0.0889 | 0.2683 | 0.3021 |
| | | TGCN | 0.1216 | 0.2572 | 0.3489 |
| | | AGCRN | 0.0977 | 0.2441 | 0.3125 |
| | | A3TGCN | 0.0768 | 0.1750 | 0.2770 |
| | Random Method | - | 0.1871 | 0.3753 | 0.4319 |
| | Time series regression Method | - | 0.1420 | 0.3539 | 0.3736 |
| **MontevideoBus** | DGSP-GCN Method | GConvGRU | 0.1216 | 0.3039 | 0.3470 |
| | | GConvLSTM | 0.0934 | 0.2595 | 0.2995 |
| | | TGCN | 0.0900 | 0.2596 | 0.3006 |
| | | AGCRN | 0.1163 | 0.2835 | 0.3341 |
| | | A3TGCN | 0.0706 | 0.2038 | 0.2717 |
| | Random Method | - | 0.1950 | 0.3685 | 0.4412 |
| | Time series regression Method | - | 0.1618 | 0.3284 | 0.4015 |
| **MetraLa** | DGSP-GCN Method | GConvGRU | 0.0456 | 0.1519 | 0.2056 |
| | | GConvLSTM | 0.0668 | 0.1755 | 0.2215 |
| | | TGCN | 0.0477 | 0.1592 | 0.1845 |
| | | AGCRN | 0.0936 | 0.2787 | 0.3150 |
| | | A3TGCN | 0.0549 | 0.1803 | 0.2343 |
| | Random Method | - | 0.1859 | 0.3625 | 0.4312 |
| | Time series regression Method | - | 0.1536 | 0.3164 | 0.3919 |

**Table 3:** The errors of the DGSP-GCN model with the A3TGCN embedding layer

| Dataset | MSE | MAE | RMSE |
|---|---|---|---|
| **Wiki Math** | 0.0983 | 0.2289 | 0.3176 |
| **Chickenpox** | 0.0502 | 0.1117 | 0.2229 |
| **PedalMe** | 0.0629 | 0.1889 | 0.2416 |
| **MontevideoBus** | 0.0607 | 0.1907 | 0.2463 |
| **MetraLa** | 0.0508 | 0.1705 | 0.2253 |

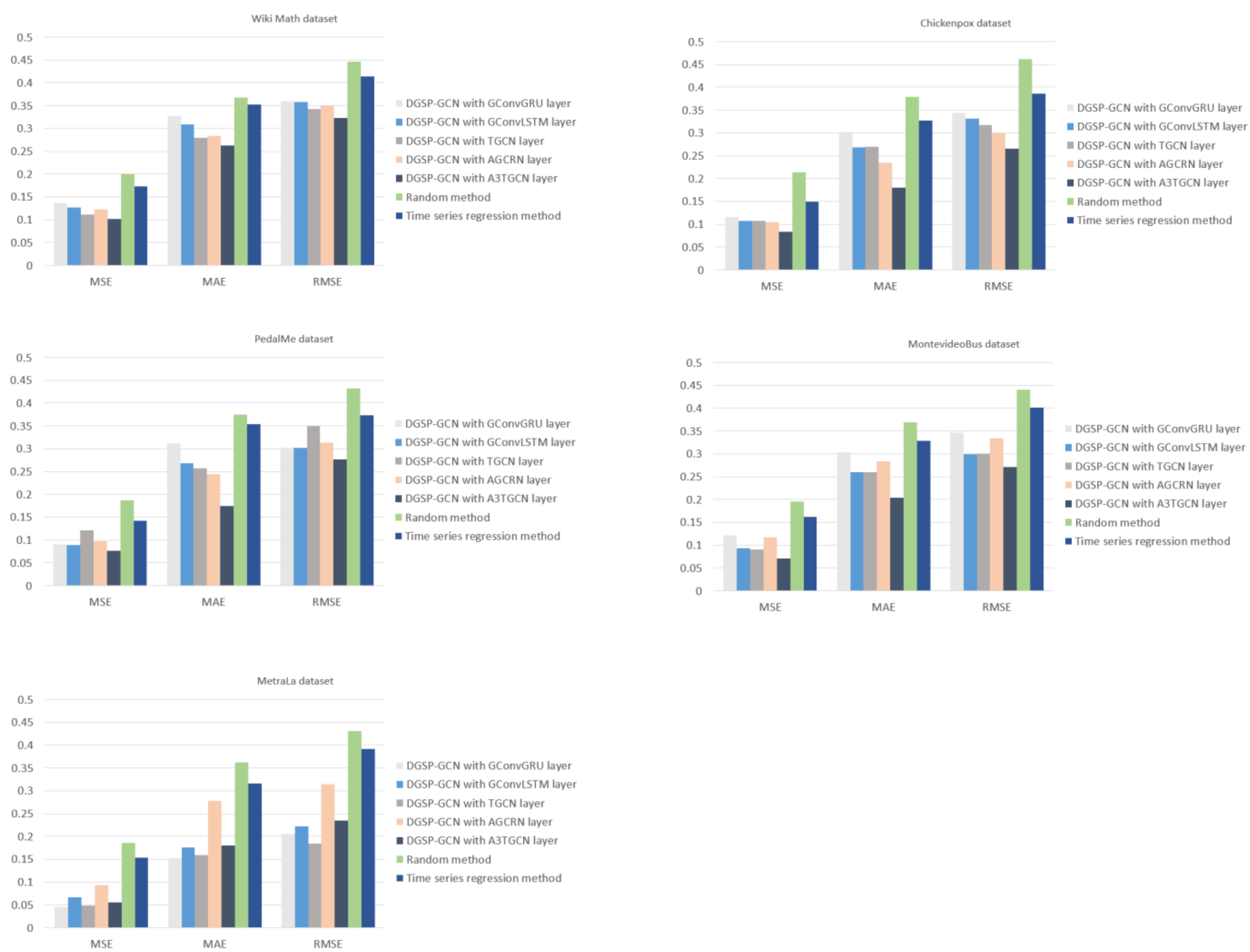

**Fig. 5.** The error rates of models for each dataset.

### 4.4. Discussion

The experimental results demonstrate that the proposed DGSP-GCN model outperforms existing methods in predicting similarity within temporal complex networks. The comparison of different embedding layers highlights the effectiveness of the A3TGCN architecture, which consistently achieves the lowest error rates across all datasets. This improvement can be attributed to the attention mechanism, which allows the model to dynamically weigh the significance of different temporal states, leading to more robust and context-aware similarity predictions.

Each embedding recurrent layer in the architectures of the proposed model has its merits and demerits, and the best layer to use will depend on our experiments. That is why, according to the results of experiments in Table 2 and Fig. 5, the performance of the proposed model in the same condition for the A3TGCN layer is almost better than others because of an attention mechanism layer. This attention-driven strategy enables A3TGCN to capture intricate relationships and the relative importance of neighboring nodes. As a result, it produces highly informative and context-aware embeddings. At its core, the attention mechanism allows the model to dynamically assign weights or importance scores to each neighbor during the aggregation process, considering both local and global information. By adaptively attending to the most relevant nodes, A3TGCN effectively focuses its attention on the crucial aspects of the graph, emphasizing nodes that contribute significantly to the target node's representation. This attention-based approach offers several significant advantages: firstly, it enables the model to assign higher weights to influential neighbors, thereby capturing the influence and impact of key nodes in the embedding process. Secondly, it allows A3TGCN to prioritize relevant structural patterns and dependencies, enhancing its ability to capture complex graph dynamics and characteristics. Thirdly, the attention mechanism enables the model to handle varying degrees of node importance, such as nodes with high centrality or rare but impactful nodes, enhancing the robustness and adaptability of the embedding generation process.

A critical aspect of this study is the comparative analysis with baseline methods, including time series regression and random similarity assignment. The results indicate that the proposed model significantly reduces error rates compared to these baselines. The time series regression method, while capable of capturing temporal trends at the node level, lacks the structural awareness necessary for graph-based similarity predictions. Consequently, it exhibits higher error rates than DGSP-GCN, particularly in datasets with complex topological dependencies. The random method serves as a lower-bound benchmark, confirming that the proposed model is meaningfully learning network dynamics rather than producing arbitrary similarity scores.

Moreover, the inclusion of multiple embedding architectures provides insights into the trade-offs between different methods. While GConvGRU, GConvLSTM, and TGCN offer competitive performance, their reliance on recurrent mechanisms without attention-based refinement limits their ability to prioritize influential nodes. The AGCRN model, which incorporates adaptive graph convolution, demonstrates notable performance improvements but still falls short of A3TGCN due to its lack of explicit temporal attention mechanisms.

One of the key findings is the adaptability of DGSP-GCN across diverse datasets, including WikiMath, Chickenpox, PedalMe, MontevideoBus, and MetraLa. The model's ability to generalize across varying network structures and temporal resolutions suggests its robustness in real-world applications. For instance, in the WikiMath dataset, where node interactions evolve daily, DGSP-GCN effectively captures the subtle changes in graph topology, resulting in lower MSE and RMSE values. Similarly, in the MetraLa dataset, which features high-frequency temporal updates, the model maintains its predictive accuracy, showcasing its scalability to different temporal granularities.

While the datasets used in this study were selected for their diversity in graph-temporal structures, we recognize the potential for extending the model’s applicability to a broader range of domains. The proposed model’s architecture is not limited to specific data types and can be applied to other dynamic data, such as social media interactions, financial transaction networks, or epidemiological data. For example, in the context of social media, nodes could represent users, and edges could represent interactions, while in financial datasets, nodes might represent accounts and edges the transactions between them. Future work could include testing the model on such datasets to validate its generalizability and adaptability across various domains, providing deeper insights into its potential applications.

In summary, the DGSP-GCN model presents a significant advancement in the evaluation of dynamic network similarity, outperforming existing approaches in accuracy and adaptability. By leveraging attention mechanisms and deep graph embeddings, it provides a more nuanced understanding of temporal graph evolution, paving the way for enhanced applications in anomaly detection, network security, and dynamic system modeling.

## 5. Conclusion

There are many different kinds of challenges in complex network modeling based on machine learning, and solving them improves the performance of network generative models, especially their dynamic counterparts. An automatic evaluation approach based on deep learning is one of the most effective ways to improve the quality of artificially produced networks. Dynamic generative models have used statistical approaches of static modeling methods, which is not optimal due to time dependency in dynamic graph-based structures. Therefore, this paper proposes a deep learning-based model to solve the challenge of evaluating dynamic generative models. The proposed model contains several phases, including node and edge embedding. In the case of embedding, we have tested several embedding architectures like GConvGRU, GConvLSTM, TGCN, AGCRN, and A3TGCN. These architectures contain

GCN and recurrent neural network layers. On the one hand, the GCN is used to capture the graph's topological structure to obtain the spatial dependence; on the other hand, the recurrent neural network layer is used to capture the dynamic change of node attribute to obtain the temporal dependence. According to the conducted tests, the A3TGCN performs almost better than other embedding layers due to having an attention mechanism layer. Besides evaluating dynamic generative models, the proposed model can also be used in anomaly detection. Our model achieved the best prediction results under different horizons when evaluated on five real-world datasets and compared with the random and time series regression baselines. In other words, according to Table 3, the average error rate of the proposed model based on MSE, MAE, and RMSE performance measures with the A3TGCN embedding layer for the datasets presented in Table 1 are equal to 0.0645, 0.1781, and 0.2507, respectively. In contrast, these averages for the time series regression model based on Table 2 each are equal to 0.1557, 0.3356, and 0.3933. Also, these values for another baseline, the random model, are separately 0.1961, 0.3623, and 0.4424. The findings of this paper demonstrate the effectiveness of the DGSP-GCN model in evaluating dynamic network generative models and detecting anomalies in temporal complex networks. By leveraging attention mechanisms and deep graph embeddings, the model provides a robust framework for capturing the evolving structural and temporal dynamics of networks, outperforming traditional methods such as time series regression and random similarity assignment. These results have significant implications for real-world applications, including network security, anomaly detection, and dynamic system modeling, where understanding the temporal evolution of networks is crucial. While the proposed method demonstrates superior performance compared to existing approaches, it is not without limitations. One key limitation lies in its reliance on temporal graphs with well-structured snapshots, which may not be readily available for all types of real-world dynamic networks. Additionally, the computational complexity of the attention mechanism in large-scale graphs may pose challenges for scalability. These issues highlight the need for further optimization of the model's architecture to reduce its computational cost and adapt to more irregular or incomplete data. In light of the findings presented in this study, there are several promising avenues for future research. It would be valuable to extend the model's capabilities to handle larger and more diverse datasets, such as real-time financial transaction networks or social media interactions, where noise and data sparsity are common challenges. Moreover, integrating adaptive learning techniques to dynamically adjust to evolving network structures could enhance the model's performance in highly dynamic and heterogeneous environments. Exploring explainability and interpretability in graph attention mechanisms is another promising direction to provide deeper insights into the decision-making process of the model. Last but not least, improving the model's ability to predict similarity for multiple future time steps ($i \geq 1$) and integrating explainability into the attention mechanisms could provide deeper insights and broader applicability in various domains.

## Statements and Declarations

**Author Contributions**

Alireza Rashnu: Conceptualization, Methodology, Software, Validation, Formal analysis, Theoretical analysis, Investigation, Data curation, Writing - original draft, Writing – review & editing, Visualization.

Sadegh Aliakbary: Supervision, Writing – review & editing, Methodology, Project administration, Theoretical analysis.

**Funding**

The authors declare that no funds, grants, or other support were received during the preparation of this manuscript.

**Competing Interests**

The authors have no relevant financial or non-financial interests to disclose.

**Data Availability**

The datasets used during the current study are publicly available via the following links:

1. WikiMath Dataset:
   https://raw.githubusercontent.com/benedekrozemberczki/pytorch_geometric_temporal/master/dataset/wikivital_mathematics.json
2. Chickenpox Dataset:
   https://raw.githubusercontent.com/benedekrozemberczki/pytorch_geometric_temporal/master/dataset/chickenpox.json
3. PedalMe Dataset:
   https://raw.githubusercontent.com/benedekrozemberczki/pytorch_geometric_temporal/master/dataset/pedalme_london.json
4. MetraLa Dataset:
   https://graphmining.ai/temporal_datasets/METR-LA.zip
5. MontevideoBus Dataset:
   https://raw.githubusercontent.com/benedekrozemberczki/pytorch_geometric_temporal/master/dataset/montevideo_bus.json